\documentclass{elsart}
\journal{New Astronomy}

\usepackage{natbib}

\usepackage{epsfig}

\usepackage{amssymb}

\def\apj{Astrophys. J.}
\def\apjs{Astrophys. J. Supp.}

\def\lsim{\mathrel{\rlap{\lower4pt\hbox{\hskip1pt$\sim$}}
    \raise1pt\hbox{$<$}}}                
\def\gsim{\mathrel{\rlap{\lower4pt\hbox{\hskip1pt$\sim$}}
    \raise1pt\hbox{$>$}}}       

\newcommand{\delT}{\Delta T}
\newcommand{\delTobs}{\widetilde{\Delta T}}

\newcommand{\lp}{{l^\prime}}
\newcommand{\Lp}{{L^\prime}}
\newcommand{\mpp}{m^{\prime\prime}}
\newcommand{\lpp}{{l^{\prime\prime}}}
\newcommand{\Mpp}{{M^{\prime\prime}}}
\newcommand{\p}{\prime}
\newcommand{\q}{\mathbf{\hat{q}}}
\newcommand{\qp}{\mathbf{\hat{q}^\prime}}

\newcommand{\cclobs}{\langle \widetilde{{C}}_l \rangle}

\def\lsim{\mathrel{\rlap{\lower4pt\hbox{\hskip1pt$\sim$}}
    \raise1pt\hbox{$<$}}}                
\def\gsim{\mathrel{\rlap{\lower4pt\hbox{\hskip1pt$\sim$}}
    \raise1pt\hbox{$>$}}}                

\begin{document}

\begin{frontmatter}



\title{ Non-Circular beam correction to \\ the CMB power spectrum}


\author{{\bf Tarun Souradeep}$^1$, Sanjit Mitra$^1$, Anand
Sengupta$^2$}
\author{Subharthi Ray$^1$, Rajib Saha$^{1,3}$ }
\address{$^1$Inter-University Centre for Astronomy and Astrophysics
(IUCAA),\\ Post Bag 4, Ganeshkhind, Pune 411~007, India.\\ E-mail:
tarun@iucaa.ernet.in; sanjit@iucaa.ernet.in; sray@iucaa.ernet.in\\$^2$
School of Physics and Astronomy, Cardiff University, \\ 5, The Parade,
Cardiff CF24 3YB, U.K.\\ E-mail: Anand.Sengupta@astro.cf.ac.uk
\\$^3$Physics Department, Indian Institute of Technology, Kanpur, \\U.P.,
208016, India.\\ E-mail: rajib@iitk.ac.in}
\begin{abstract}
In the era of high precision CMB measurements, systematic effects are
beginning to limit the ability to extract subtler cosmological
information.  The non-circularity of the experimental beam has become
progressively important as CMB experiments strive to attain higher
angular resolution and sensitivity. The effect of non-circular beam on
the power spectrum is important at multipoles larger than the
beam-width.  For recent experiments with high angular resolution,
optimal methods of power spectrum estimation are computationally
prohibitive and sub-optimal approaches, such as the Pseudo-$C_l$
method, are used. We provide an analytic framework for correcting the
power spectrum for the effect of beam non-circularity and non-uniform
sky coverage (including incomplete/masked sky maps). The approach is
perturbative in the distortion of the beam from non-circularity
allowing for rapid computations when the beam is mildly
non-circular. When non-circular beam effect is important, we advocate
that it is computationally advantageous to employ `soft' azimuthally
apodized masks whose spherical harmonic transform die down fast with
$m$.
\end{abstract}


\end{frontmatter}

\section{Introduction}

The fluctuations in the Cosmic Microwave background (CMB) radiation
are theoretically very well understood, allowing precise and
unambiguous predictions for a given cosmological
model~\cite{bonLH,hu_dod02}. Consequently, measurement of CMB
anisotropy has spearheaded the remarkable transition of cosmology into
a precision science. The transition has also seen the emergence of
data analysis of large complex data sets as an important and
challenging component of research in cosmology. Increasingly
sensitive, high resolution, measurements over large regions of the sky
pose a stiff challenge for current analysis techniques to realize the
full potential of precise determination of cosmological
parameters. The analysis techniques must not only be computationally
fast to contend with the huge size of the data, but, the higher
sensitivity also limits the simplifying assumptions that can be then
invoked to achieve the desired speed without compromising the final
precision goals. There is a worldwide effort to push the boundary of
this inherent compromise faced by the current CMB experiments that
measure the anisotropy in the CMB temperature and its polarization.

Accurate estimation of the angular power spectrum, $C_{l}$, is
arguably the foremost concern of most CMB experiments. The extensive
literature on this topic has been summarized in
literature~\cite{hu_dod02,efs04}. For Gaussian, statistically
isotropic CMB sky, the $C_l$ that corresponds to the covariance that
maximizes the multivariate Gaussian PDF of the temperature map, $\Delta
T(\q)$ is the Maximum Likelihood (ML) solution. Different ML
estimators have been proposed and implemented on CMB data of small and
modest sizes~\cite{gor94,gor_hin94,gor96,gor97,max97,bjk98}. While it
is desirable to use optimal estimators of $C_l$ that obtain (or
iterate toward) the ML solution for the given data, these methods are
usually limited by the computational expense of matrix inversion that
scales as $N_d^3$ with data size $N_d$~\cite{bor99,bon99}.  Various
strategies for speeding up ML estimation have been proposed, such as,
exploiting the symmetries of the scan strategy~\cite{oh_sper99}, using
hierarchical decomposition~\cite{dor_knox01}, iterative multi-grid
method~\cite{pen03}, etc. Variants employing linear combinations of
$\Delta T(\q)$ such as $a_{lm}$ on set of rings in the sky can
alleviate the computational demands in special
cases~\cite{harm02,wan_han03}.  Other promising `exact' power
estimation methods have been recently
proposed~\cite{knox01,wan04,jew04}.

However there also exist computationally rapid, sub-optimal estimators
of $C_l$. Exploiting the fast spherical harmonic transform ($\sim
N_d^{3/2}$), it is possible to estimate the angular power spectrum
$C_l= \sum_m |a_{lm}|^2/(2l+1)$ rapidly~\cite{yu_peeb69,peeb73}. This
is commonly referred to as the Pseudo-$C_l$
method~\cite{wan_hiv03}. \footnote{Analogous approach employing fast
estimation of the correlation function $C(\q\cdot\qp)$ have also been
explored~\cite{szap01,szap_prun01}.}  It has been recently argued that
the need for optimal estimators may have been over-emphasized since
they are computationally prohibitive at large $l$ . Sub-optimal
estimators are computationally tractable and tend to be nearly optimal
in the relevant high $l$ regime. Moreover, already the data size of
the current sensitive, high resolution, `full sky' CMB experiments
such as WMAP have been compelled to use sub-optimal Pseudo-$C_l$
related methods~\cite{Bennett:2003, hin_wmap03}.  On the other hand,
optimal ML estimators can readily incorporate and account for various
systematic effects, such as noise correlations, non-uniform sky
coverage and beam asymmetries. The systematic correction to the
Pseudo-$C_l$ power spectrum estimate arising from non-uniform sky
coverage has been studied and implemented for CMB
temperature~\cite{master} and polarization~\cite{Brown:2004}.  The
systematic correction for non circular beam has been studied by
us~\cite{mit04}. Here we extend the results to include non-uniform sky
coverage.

It has been usual in CMB data analysis to assume the experimental beam
response to be circularly symmetric around the pointing
direction. However, any real beam response function has deviations
from circular symmetry. Even the main lobes of the beam response of
experiments are generically non-circular (non-axisymmetric) since
detectors have to be placed off-axis on the focal plane. (Side lobes
and stray light contamination add to the breakdown of this
assumption). For highly sensitive experiments, the systematic errors
arising from the beam non-circularity become progressively more
important.  Dropping the circular beam assumption leads to  major
complications at every stage of analysis pipeline. The extent to which
the non-circularity affects the step of going from the time-stream
data to sky map is very sensitive to the scan-strategy. The beam now
has an orientation with respect to the scan path that can potentially
vary along the path. This implies that the beam function is inherently
time dependent and difficult to deconvolve. 

\begin{figure}
\centering
\includegraphics[angle = 0,scale = 0.45 ]{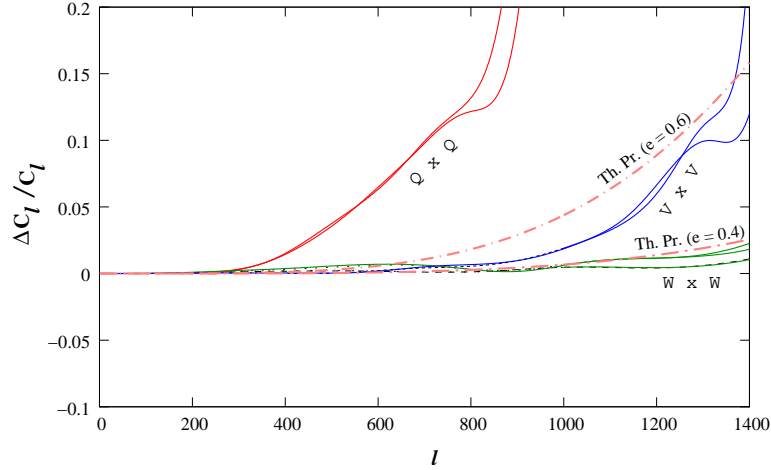}
\caption{The predicted non-circular beam correction for a CMB
experiment with elliptical Gaussian beam with {\it fwhm} beam-width of
$0.22^\circ$ and two values of eccentricity, $e=0.6$ and $e=0.4$ are
shown as dashed-dot lines (labeled Th. Pr.). The solid curves are the
non circular beam corrections estimated by the WMAP team for the Q,V
and W channels. The {\it fwhm} beam-width of $0.22^\circ$ corresponds
closely to the cleanest and highest resolution W-band beam-width. For
larger beam-widths, the theoretical curves would roughly shift left in
multipole by the ratio of beam-widths.}
\label{clerr}
\end{figure}

Even after a sky map is made, the non-circularity of the effective
beam affects the estimation of the angular power spectrum, $C_l$, by
coupling the power at different multipoles, typically, on scales
beyond the inverse angular beam-width.  Mild deviations from
circularity can be addressed by a perturbation
approach~\cite{TR:2001,fos02} and the effect of non-circularity on the
estimation of CMB power spectrum can be studied (semi)
analytically~\cite{mit04}.  Fig.~\ref{clerr} shows the predicted level
of non-circular beam correction in our formalism for elliptical beams
with {\it fwhm} beam-width of $0.22^\circ$ compared to the
non-circular beam corrections computed in the recent data release by
WMAP~\cite{hin_wmap06}.

To avoid contamination of the primordial CMB signal by Galactic
emission, the region adjoining the Galactic plane is masked from
maps. If the Galactic cut is small enough, then the coupling matrix
will be invertible, and the two-point correlation function can be
determined on all angular scales from the data within the uncut
sky~\cite{mch02}.  Hivon et al.~\cite{master} present a technique
(MASTER) for fast computation of the power spectrum taking accounting
for the galactic cut, but for circular beams. {\em In our present
work, we present analytical expressions for the bias matrix of the
Pseudo-${C_l}$ estimator for the incomplete sky coverage, using a
non-circular beam.}

\section{CMB angular Power spectrum estimation: \hfil\break Non-circular beam 
\& Non-uniform coverage effects}

The observed CMB temperature fluctuation is convolved with a beam
function and contaminated by noise. Further, the CMB signal cannot be
obtained for full sky because of galactic contamination (and
extragalactic point sources). We derive the general form of the bias
matrix including non-uniform/incomplete sky coverage and a general
beam function.
 
The observed temperature fluctuation field $\delTobs(\q)$ is the
convolution of the ``beam" profile $B(\q,\qp)$ with the real
temperature fluctuation field $\delT(\q)$ (ignoring the additive noise
term for simplicity)~:

\begin{equation}
\delTobs(\q) \ = \ \int_{4\pi} d\Omega_{\qp} \, B(\q,\qp) \,
\delT(\qp).
\end{equation}

The two point correlation function for a statistically isotropic CMB
anisotropy signal is
\begin{equation}
C(\q,\q^\prime) =  \langle \delTobs(\q) \delTobs(\qp) \rangle =
   \sum^\infty_{l = 0} {(2 l + 1) \over 4\pi}\, C_l\,\,
   W_l(\q,\,\q^\prime )\,,
   \label{corre}
\end{equation}
where $C_l$ is the angular spectrum of CMB anisotropy signal and
the window function
\begin{equation}
W_l(\q_1,\q_2) \ \equiv \ \int d\Omega_\q \int d\Omega_\qp \,
B(\q_1, \q) B(\q_2, \qp) P_l(\q \cdot \qp), \label{eq:defW}
\label{windef}
\end{equation}
encodes the effect of finite resolution through the beam function.
A CMB anisotropy experiment probes a range of angular
scales characterized by a \textit{window} function
$W_l(\q,\q^\prime)$.  The window depends both on the scanning
strategy as well as the angular resolution and response of the
experiment. However, it is neater to logically separate these two
effects by expressing the window $W_l(\q,\q^\prime)$ as a sum of
`elementary' window function of the CMB anisotropy at each point
of the map~\cite{TR:2001}. For a given scanning
strategy, the results can be readily generalized using the
representation of the window function as sum over elementary
window functions (see, {\it e.g.,} \cite{TR:2001,pyV}). 

For some experiments, the beam function may be assumed to be
circularly symmetric about the pointing direction, i.e., $B(\q,
\q^\prime) \equiv B(\q\cdot\q^\prime)$ without significantly
affecting the results of the analysis.  In any case, this
assumption allows a great simplification since the beam function
can then be represented by an expansion in Legendre polynomials as
\begin{equation}
B(\q\cdot\qp) \ = \ \frac{1}{4\pi}\,\sum_{l=0}^\infty\, (2l+1)\,
B_l\, P_l(\q\cdot\qp). \label{eq:BthetafromBl}
\end{equation}
Consequently, it is straightforward to derive the well known
simple expression
\begin{equation}
   W_l(\q,\,\qp) \ = \ B^2_l \, P_l(\q\cdot\qp)\,,
   \label{isowine}
\end{equation}
for a circularly symmetric beam function.

We define the Pseudo-$C_l$ estimator as
\begin{equation}
\widetilde{{C}}_l \ \equiv \ \frac{1}{4\pi}
\int_{4\pi} d\Omega_{\q_1} \int_{4\pi} d\Omega_{\q_2} \, U(\q_1)
\, U(\q_2) \, P_l(\q_1 \cdot \q_2) \, \delTobs(\q_1) \,
\delTobs(\q_2) ,
\end{equation}
where $U(\q)$ denotes the mask function representing the incomplete
sky.  The expectation value of the Pseudo-$C_l$ estimator can be shown
to take the form
\begin{eqnarray}
\!\!\!\cclobs = \frac{1}{2l+1} \sum_{\lp=0}^{\infty}
  {C}_\lp \sum_{n=-l}^l\!
\sum_{m=-\lp}^\lp \!\left|\int_{4\pi}\!\!\! d\Omega_{\q} U(\q) \, Y_{l
n}(\q) \left[ \int_{4\pi}\!\!\! d\Omega_{\qp} \, Y^*_{\lp m}(\qp) \,
B(\q,\qp) \right] \right|^2.
\label{eq:pscl1}
\end{eqnarray}
The integral in the square bracket can be simplified to~\cite{TR:2001}
\begin{equation}
\int_{4\pi} d\Omega_{\qp} \, Y^*_{\lp m}(\qp) \, B(\q,\qp)  \ = \
\sqrt{\frac{2\lp+1}{4\pi}} \sum_{m^\p=-\lp}^\lp B_\lp \,
\beta_{\lp m^\p} \, D^\lp_{mm^\p}(\q,\rho(\q)).
\end{equation}
The Beam Distortion Parameter (BDP) $\beta_{lm} \equiv
b_{lm}/b_{l0}$ is expressed in terms of

\begin{eqnarray}
b_{lm}  \equiv \int_{4\pi} d\Omega_{\mathbf{\hat{q}}} \,
Y_{lm}^*({\mathbf{\hat{q}}}) \, B(\mathbf{\hat{z}},\mathbf{\hat{q}})\,,
\nonumber\\
B_l  \equiv \int_{-1}^{1} d(\q \cdot \qp) \, P_l(\q \cdot
\qp)\, \mathcal{B}(\q \cdot \qp) \label{bdp}
\end{eqnarray}

where $\mathcal{B}(\q \cdot \qp)$ is the \textit{circularized}
beam obtained by averaging $B(\mathbf{\hat{z}},\mathbf{\hat{q}})$
over azimuth $\phi$. Hence,

\begin{equation}
\nonumber
B_l  = \int_{0}^{\pi} \sin\theta d\theta \,
\sqrt{\frac{4\pi}{2l+1}} \, Y_{l0}^*({\mathbf{\hat{q}}}) \left[
\frac{1}{2\pi} \int_{0}^{2\pi} d\phi \,
B(\mathbf{\hat{z}},\mathbf{\hat{q}}) \right]  
= \sqrt{\frac{4\pi}{2l+1}} \, b_{l0}. \label{Blbl0}
\end{equation}

Making a spherical harmonic expansion of the
mask function $U(\q)$
\begin{equation}
U(\q) \ = \ \sum_{l=0}^{\infty} \sum_{m=-l}^{l} U_{lm} \,
Y_{lm}(\q) \label{maskTr}
\end{equation}
we can simplify Eq.\ref{eq:pscl1} as
\begin{equation}
\cclobs \ = \ \sum_{\lp} A_{l\lp} C_\lp\,.
\end{equation}
The general form of the \emph{bias matrix}, $A_{l\lp}$ is thus given
by

\begin{equation}
A_{l l^\prime} \ = \ \frac{B_l^2}{4\pi}
\frac{(2l^\prime+1)}{(2l+1)} 
\sum_{n=-l}^{l} \sum_{m=-l^\prime}^{l^\prime}
\left|\sum_{m^\prime=-l^\prime}^{l^\prime} \beta_{l^\prime
m^\prime} \sum_{\lpp =0}^{\infty} \sum_{\mpp =-\lpp}^{\lpp}
U_{\lpp\mpp} \, J^{l\lpp\lp}_{n\mpp mm^\prime} \right|^2,
\label{gen}
\end{equation}
where
\begin{equation}
J^{l\lpp\lp}_{n\mpp mm^\prime} \ \equiv \ \int_{4\pi}
d\Omega_{\mathbf{\hat{q}}} \, Y_{ln}(\mathbf{\hat{q}}) \,
Y_{\lpp\mpp}(\mathbf{\hat{q}}) \,
D_{mm^\prime}^{l^\prime}(\mathbf{\hat{q}},\rho(\mathbf{\hat{q}})).
\label{defJ}
\end{equation}

To proceed further \emph{analytically}, we need a model for
$\rho(\q)$. We shall continue assuming \emph{non-rotating} beams,
i.e. $\rho(\q) = 0$.  We evaluate the integral $J^{l\lpp\lp}_{n\mpp
mm^\prime}$, with two different approaches.  In the first method,
using only the sinusoidal expansion of Wigner-$d$, we get
\begin{eqnarray}
J^{l\lpp\lp}_{n\mpp mm^\prime} &=&  2\pi \, \delta_{\mpp (m-n)}
\frac{\sqrt{(2l+1)(2\lpp+1)}}{4\pi} \sum_{M=-l}^{l}
d^{l}_{nM}\left(\frac{\pi}{2}\right)
d^{l}_{M0}\left(\frac{\pi}{2}\right) \nonumber\\&&{}\times\sum_{\Mpp=-\lpp}^{\lpp}
d^{\lpp}_{\mpp\Mpp}   \left(\frac{\pi}{2}\right)
d^{\lpp}_{\Mpp0}\left(\frac{\pi}{2}\right)
 \sum_{M^\p=-\lp}^{\lp}
d^{\lp}_{mM^\p}\left(\frac{\pi}{2}\right) d^{\lp}_{M^\p
m^\p}\left(\frac{\pi}{2}\right) \nonumber\\&&{}\times\left[ i^{n+m+m^\prime+\mpp} \,
(-1)^{M+\Mpp+M^\p} \int_0^\pi \sin\theta d\theta \,
e^{i(M+M^\p+\Mpp)\theta} \right].
\end{eqnarray}
In the alternative method using Clebsch Gordon coefficients, we can
evaluate $J^{l\lpp\lp}_{n\mpp mm^\prime}$ as:
\begin{eqnarray}
J^{l\lpp\lp}_{n\mpp mm^\prime}&&= (-1)^{n+\mpp} \,
\delta_{\mpp(m-n)} \, \frac{\sqrt{(2l+1)(2\lpp+1)}}{2}\nonumber \\
&& \times\sum_{L=|l-\lpp|}^{l+\lpp} C^{L0}_{l0\lpp 0} 
C^{L(n+\mpp)}_{ln\lpp \mpp} \sum_{\Lp=|L-\lp|}^{L+\lp}
C^{\Lp(m-n-\mpp)}_{L(-n-\mpp)\lp m} \, C^{\Lp m^\p}_{L0\lp m^\p}
\nonumber \\
&&\times\sum_{N=-\Lp}^{\Lp} \, d^{\Lp}_{0N}\left(\frac{\pi}{2}\right) \,
d^{\Lp}_{Nm^\p}\left(\frac{\pi}{2}\right) \,\left[ i^{m^\p} \,
(-1)^{N} \int_0^\pi \sin\theta d\theta \,
e^{i N\theta} \right].
\end{eqnarray}
The analytic expressions reduce to the known analytical results for
circular beam and non-uniform sky coverage studied
in~ref.~\cite{master} and our earlier results for non-circular beam
for full sky~\cite{mit04}.

These results offer the possibility of rapid estimation of the
non-circular beam effect in the Pseudo-$C_l$ estimation. The
expression in terms of the $d_{mm'}^l(\pi/2)$ coefficients is the
computationally superior approach. These coefficients can be computed
using stable recurrence relations~\cite{ris96,wan_gor01}.  In a more
detailed publication we describe the algorithm in more
detail~\cite{usinprep}. The expressions also highlight the two aspects
to speeding up the computation of the systematic effect:
\begin{itemize}
\item[i.] Mildly non-circular beams where the beam distortion
parameters (BDP), $\beta_{lm}$ at each $l$ fall off rapidly with
$m$. This allows us to neglect $\beta_{lm}$ for $m>m_{\mathrm
beam}$. For most real beams, $m_{\mathrm{beam}} \sim 4$ is a
sufficiently good approximation.  This cuts-off the summations over
BDP in the expressions for $A_{l\lp}$.
\item[ii.] Soft, azimuthally apodized, masks where the coefficients
$U_{lm}$ are small beyond $m>m_{\mathrm{mask}}$. Moreover, it is
useful to smooth the mask in $l$, such the $U_{lm}$ die off rapidly
for $l>l_{\mathrm{mask}}$ too.
\end{itemize}
The mild-circularity perturbation approach has been introduced and
discussed in ref.~\cite{TR:2001}. The circularity of the beam has to
be addressed in the design of the CMB experiments. Our results suggest
the systematics due to non-circular distortions of the beam are
manageable if one ensures the large BDP are limited to a few $m$
(i.e., narrow band limited violation of axis-symmetry). The beams for
many experiments, such as Python-V and WMAP are well approximated as
elliptical Gaussian functions~\cite{TR:2001,pyV,mit04}.  For radically
non-axisymmetric beams, modeling the beam in terms of superposition of
displaced circular Gaussian beams has been
proposed~\cite{asym_04,xspect_05}. Our approach allows a simple, cost
effective extension to modeling with the more general elliptical
Gaussian beams, or other mildly non-circular beam forms.

\begin{figure}
\centering
\includegraphics[angle = 90 ,scale = 0.45 ]{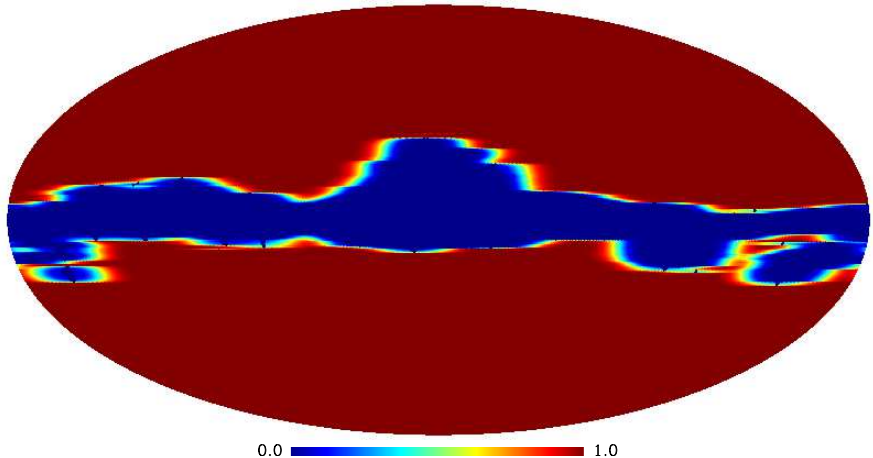}
\caption{A softened (azimuthally smoothed) mask reconstructed from the
WMAP Kp2 mask which would reduce the computational cost of estimating
the non-circular beam effect on the angular power spectrum.}
\label{reconmask}
\end{figure}
The mask of the galactic region $U(\q)$ can be chosen at the time of
data analysis. The coupling of BDP with $U_{lm}$ suggests that a
judicious choice of mask reduces the computational costs of
non-circular beam corrections. Fig.~\ref{reconmask} shows a softened
version of the Kp2 mask used by the WMAP team~\cite{Bennett1}, where
the mask is azimuthally smoothed. The final apodized mask is obtained
by multiplying an azimuthally smoothed mask raised to a sufficiently
large power with the original mask and has reduced power at large $m$
(i.e., $U_{lm}$ is negligible for
$m>m_{\mathrm{mask}}$)~\footnote{Recently, apodized masks (using
circularly symmetric smoothing kernels) have been recommended in the
context of CMB polarization maps~\cite{efs06}. In our context, the
mask is predominantly azimuthally smoothed to retain more CMB sky.}.
In a forthcoming publication we describe the method of making soft
masks~\cite{usinprep}.  For mildly non-circular, nearly azimuthally
symmetric case, the required number of computation cycle to compute
the bias matrix up to a multipole $l$ scales as $\sim
(2m_{\mathrm{beam}}+1) (2m_{\mathrm{mask}}+1) \, l^5$ up to leading
order for $l \gg l_{\mathrm{mask}}$. Here, $m_{\mathrm{beam}}$ is the
cut-off in the beam distortion parameters (BDP), $\beta_{lm}$ and
$m_{\mathrm{mask}}$ is the cut-off in $U_{lm}$.

\section{Discussion \& Conclusion}
\label{disc}

The assumptions of non-circular beam leads to major complications at
every stage of the data analysis pipeline.  The extent to which the
non-circularity affects the step of going from the time-stream data to
sky map is very sensitive to the scan-strategy. The beam now has an
orientation with respect to the scan path that can potentially vary
along the path. This implies that the beam function is inherently time
dependent and difficult to deconvolve.

We extend our analytic approach for addressing the effect of
non-circular experimental beam function in the estimation of the
angular power spectrum $\mathcal{C}_l$ of CMB anisotropy, which also
includes the effect of the galactic cut in the entire sky map.
Non-circular beam effects can be modeled into the covariance functions
in approaches related to maximum likelihood
estimation~\cite{max97,bjk98} and can also be included in the Harmonic
ring~\cite{harm02} and ring-torus estimators~\cite{wan_han03}.
However, all these methods are computationally prohibitive for high
resolution maps and, at present, the computationally economical
approach of using a Pseudo-$C_l$ estimator appears to be a viable
option for extracting the power spectrum at high
multipoles~\cite{efs04}. The Pseudo-$C_l$ estimates have to be
corrected for the systematic biases.  While considerable attention has
been devoted to the effects of incomplete/non-uniform sky coverage, no
comprehensive or systematic approach is available for non-circular
beam.  The high sensitivity, `full' (large) sky observation from space
(long duration balloon) missions have alleviated the effect of
incomplete sky coverage and other systematic effects such as the one
we consider here have gained more significance. Non-uniform coverage,
in particular, the galactic masks affect only CMB power estimation at
the low multipoles. The analysis accompanying the recent second data
from WMAP uses the hybrid strategy~\cite{efs04} where the power
spectrum at low multipoles is estimated using optimal Maximum
Likelihood methods and Pseudo-$C_l$ are used for large
multipoles~\cite{hin_wmap06,sperg_wmap06}.

The non-circular beam is an effect that dominates at large $l$
comparable to the inverse beam width~\cite{mit04}. For high resolution
experiments, the optimal maximum likelihood methods which can account
for non-circular beam functions are computationally prohibitive.  In
implementing the Pseudo-$C_l$ estimation, we have included both the
non-circular beam effect and the effect of non-uniform sky
coverage. Our work provides a convenient approach for estimating the
magnitude of these effects in terms of the leading order deviations
from a circular beam and azimuthally symmetric mask. The perturbation
approach is very efficient. For most CMB experiments the leading few
orders capture most of the effect of beam
non-circularity~\cite{TR:2001}. Our results highlight the advantage of
azimuthally smoothed masks (mild deviations from azimuthal symmetry)
in reducing computational costs.  The numerical implementation of our
method can readily accommodate the case when pixels are revisited by
the beam with different orientations. Evaluating the realistic bias
and error-covariance for a specific CMB experiment with non-circular
beams would require numerical evaluation of the general expressions
for $A_{l\lp}$ using real scan strategy and account for inhomogeneous
noise and sky coverage, the latter part of which has been addressed in
this present work.

It is worthwhile to note in passing that that the angular power $C_l$
contains all the information of Gaussian CMB anisotropy only under the
assumption of statistical isotropy.  Gaussian CMB anisotropy map
measured with a non-circular beam corresponds to an underlying
correlation function that violates statistical isotropy. In this case,
the extra information present may be measurable using, for example,
the bipolar power
spectrum~\cite{haj_sour03,haj_sour05,haj_sour06,bas06}. Even when the
beam is circular the scanning pattern itself is expected to cause a
breakdown of statistical isotropy of the measured CMB anisotropy
~\cite{master}. For a non-circular beam, this effect could be much
more pronounced and, perhaps, presents an interesting avenue of future
study.

In addition to temperature fluctuations, the CMB photons coming from
different directions have a random, linear polarization. The
polarization of CMB can be decomposed into $E$ part with even parity
and $B$ part with odd parity.  Besides the angular spectrum
$C_l^{TT}$, the CMB polarization provides three additional spectra,
$C_l^{TE}$, $C_l^{EE}$ and $C_l^{BB}$ which are invariant under parity
transformations. The level of polarization of the CMB being about a
tenth of the temperature fluctuation, it is only very recently that
the angular power spectrum of CMB polarization field has been
detected. The Degree Angular Scale Interferometer (DASI) has measured
the CMB polarization spectrum over limited band of angular scales in
late 2002~\cite{kov_dasi02}. The DASI experiment recently published
3-year results of much refined measurements~\cite{dasi_3y}. More
recently, the BOOMERanG collaboration reported new measurements of CMB
anisotropy and polarization spectra~\cite{boom_polar}. The WMAP mission
has also measured CMB polarization spectra
~\cite{kog_wmap03,pag_wmap06}. Correcting for the systematic effects
of a non-circular beam for the polarization spectra is expected to
become important. Extending this work to the case CMB polarization is
another line of activity we plan to undertake in the near future.

In summary, we have presented a perturbation framework to compute the
effect of non-circular beam function on the estimation of power
spectrum of CMB anisotropy taking into account the effect of a
non-uniform sky coverage (eg., galactic mask). We not only present the
most general expression including non-uniform sky coverage as well as
a non-circular beam that can be numerically evaluated but also provide
elegant analytic results in interesting limits.  As CMB experiments
strive to measure the angular power spectrum with increasing accuracy
and resolution, the work provides a stepping stone to address a rather
complicated systematic effect of non-circular beam functions.

\section*{Acknowledgments}

We thank Olivier Dore and Mike Nolta for providing us with the data
files of the non-circular beam correction estimated by the WMAP team.
We thank Kris Gorski, Jeff Jewel \& Ben Wandelt for the reference and
providing a code for computing Wigner-$d$ functions. We have benefited
from discussions with Francois Bouchet and Simon Prunet. Computations
were carried out at the HPC facility of IUCAA.

\end{document}